\documentclass[twocolumn,showpacs,amsmath,amssymb,floatfix,superscriptaddress]{revtex4}
\usepackage{graphicx}
\usepackage{dcolumn}
\usepackage{bm}
\usepackage{hyperref} 
\usepackage{latexsym}

\begin{document}
\title{Constrained Opinion Dynamics: Freezing and Slow Evolution}  
\author{F. Vazquez}\email{fvazquez@buphy.bu.edu}
\author{P.~L.~Krapivsky}\email{paulk@bu.edu}
\author{S.~Redner}\email{redner@bu.edu}
\affiliation{Center for BioDynamics, Center for Polymer Studies, 
and Department of Physics, Boston University, Boston, MA, 02215}

\begin{abstract}
  
  We study opinion formation in a population of leftists, centrists, and
  rightist.  In an interaction between neighboring agents, a centrist and a
  leftist can become both centrists or leftists (and similarly for a centrist
  and a rightist), while leftists and rightists do not affect each other.
  For any spatial dimension the final state is either consensus (of one of
  three possible opinions), or a frozen population of leftists and rightists.
  In one dimension, the opinion evolution is mapped onto a constrained spin-1
  Ising model with zero-temperature Glauber kinetics.  The approach to the
  final state is governed by a $t^{-\psi}$ long-time tail, with $\psi$ a
  non-universal exponent that depends on the initial densities.  In the
  frozen state, the length distribution of single-opinion domains has an
  algebraic small-size tail $x^{-2(1-\psi)}$ with average domain length
  $L^{2\psi}$, where $L$ is the length of the system.

\end{abstract}  


\maketitle
  
One of the basic issues in opinion dynamics is to understand the conditions
under which consensus or diversity is reached from an initial population of
individuals (agents) with different opinions.  Models for such evolution are
typically based on each agent freely adopting a new state in response to
opinions in a local neighborhood \cite{F74}.  The attribute of
incompatibility -- in which agents with sufficiently disparate opinions do
not interact -- has recently been found to prevent ultimate consensus from
being reached \cite{LN97,opinion}.  Related phenomenology arises in the
Axelrod model \cite{axelrod, CMV}, a simple model for the formation and
evolution of cultural domains.  The goal of the present paper is to
investigate the role of incompatibility within a minimal model for opinion
dynamics.  This constraint has a profound effect on the nature of the final
state.  Moreover, there is anomalously slow relaxation to the final state.
While we primarily frame our discussion in terms of opinion dynamics, our
results apply equally well to the coarsening of spin systems.  In the latter
context, we obtain a new non-universal kinetic exponent in one dimension that
originates from topological constraints on the arrangement of spins.

\begin{figure}[ht] 
 \vspace*{0.cm}\hspace*{0.1in}
 \includegraphics*[width=0.375\textwidth]{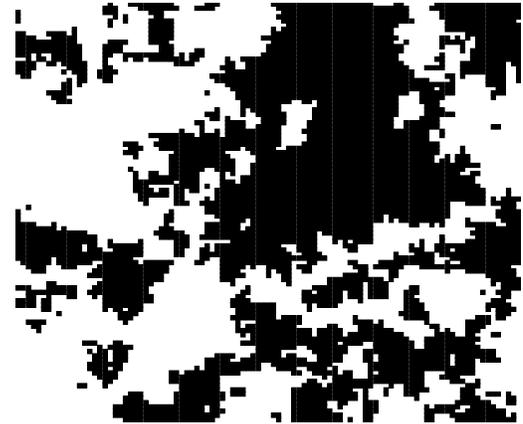}
\caption{Typical frozen final state in our opinion dynamics
  model on a $100\times 100$ square lattice for $\rho_0=0.1$.  The two
  extreme opinions are represented by black and white squares.}
\label{map}
\end{figure}

We consider a ternary system in which each agent can adopt the opinions of
leftist, centrist, and rightist.  The agents populate a lattice and in a
single microscopic event an agent adopts the opinion of a randomly-chosen
neighbor, but with the crucial proviso that that leftists and rightists are
considered to be so incompatible that they do not interact.  While a leftist
cannot directly become a rightist (and vice versa) in a single step, the
indirect evolution leftist\, $\Rightarrow$\, centrist\, $\Rightarrow$\,
rightist is possible.  Our model is similar to the classical voter model
\cite{voter} and also turns out to be isomorphic to the 2-trait 2-state
Axelrod model \cite{axelrod, CMV}.  Due to the incompatibility constraint in
our model, the final opinion outcome can be either consensus or a frozen
mixture of extremists with no centrists.  Figure ~\ref{map} shows a typical
frozen state on the square lattice (with periodic boundary conditions).
Notice the nested enclaves of opposite opinions and the clearly visible
clustering.

We can exploit the connection between the voter model and our opinion
dynamics model to infer the final state of the latter.  If we temporarily
disregard the difference between leftists and rightists, the resulting binary
system of centrists and extremists reduces to the voter model, for which one
of two absorbing states --- either all centrists or all extremists --- is
eventually reached.  In the context of the ternary opinion system, the latter
event can mean either a consensus of extremists or a frozen mixed state of
leftists and rightists, as depicted in Fig.~\ref{map}.  

\begin{figure}[ht] 
 \vspace*{0.cm} \includegraphics*[width=0.4\textwidth]{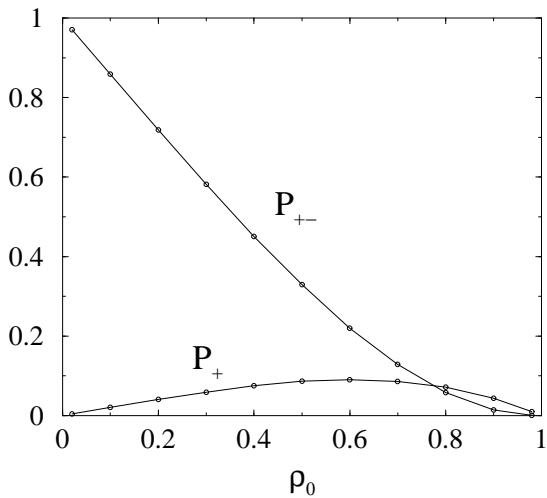}
\caption{Probability for the occurrence of a given final state as a function
  of $\rho_0$ for $\rho_+=\rho_-$.  Here $P_+$ is the probability that $+$
  consensus is reached and $P_{+-}$ is the probability that the final state
  is a frozen mixture of $+$ and $-$. }
\label{final-state}
\end{figure}

Because of the underlying voter model dynamics, the average density of each
species is globally conserved in any spatial dimension.  Therefore $\langle
\rho_i(t)\rangle =\rho_i(t=0)$, where $i$ refers to one of the states
$(+,0,-)$ and the angle brackets denote an average over all realizations of
the dynamics and over all initial states that are compatible with the
specified densities.  As a result of this conservation law, with probability
$P_0=\rho_0$ the final state consists of all centrists and with probability
$1-\rho_0$ there are no centrists in the final state.  In the latter case,
there can be either a consensus of $+$ (this occurs with probability $P_+$),
consensus of $-$ (probability $P_-$), or a frozen mixed state (probability
$P_{+-}$).  Figure \ref{final-state} shows the dependence of these final
state probabilities on $\rho_0$ in the mean-field limit (where all agents are
interconnected) in the symmetric case $\rho_+=\rho_-=(1-\rho_0)/2$.  In one
and two dimensions, the final state probabilities are nearly identical to the
mean-field predictions when $\rho_+=\rho_-$, but differences become apparent
in the strongly asymmetric cases of $\rho_+\gg \rho_-$.  The final state
probabilities also depend very weakly on the system size.

We now focus on the one dimensional case.  Here our opinion dynamics model is
equivalent to a constrained spin-1 Ising chain that is endowed with
single-spin flip zero-temperature Glauber kinetics \cite{glauber}, with
leftist, centrist, and rightist opinions equivalent to the spin states $-$,
0, and $+$, respectively.  The incompatibility constraint means that
neighboring $+$ and $-$ spins do not interact.

\begin{figure}[ht] 
 \vspace*{0.cm}\hspace*{0.1in}
 \includegraphics*[width=0.36\textwidth]{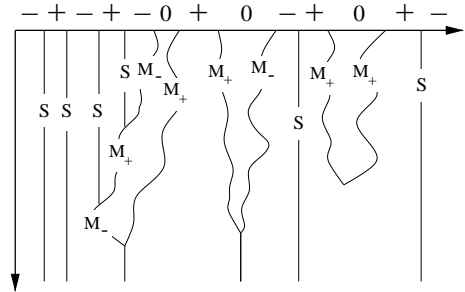}
\caption{Space-time representation of the domain wall dynamics.  Time
  runs vertically downward.  The spin state of the domains and the identity
  of each domain wall are indicated.}
\label{dw}
\end{figure}

This Ising model picture suggests that the best way to analyze the dynamics
in one dimension is to reformulate the system in terms of domain walls.
There are three types of domain walls: freely diffusing mobile domain walls
between $+0$ and between $-0$, denoted by $M_+$ and $M_-$, respectively, and
stationary domain walls $S$ between $+-$.  The mobile walls evolve by
\begin{eqnarray}
\label{MM}
M_\pm+M_\pm \to \emptyset, \qquad
M_\pm+M_\mp\to S.
\end{eqnarray}  
When a mobile wall hits a stationary wall, the former changes its sign while
the latter is eliminated via the reaction
\begin{eqnarray}
\label{MSM}
&&M_\pm+S\to M_\mp.
\end{eqnarray}  
Thus stationary domain walls are dynamically invisible; their only effect is
that the sign of a mobile wall changes whenever it meets a stationary wall,
after which the latter disappears (Fig.~\ref{dw}).  The inertness of the
stationary walls is reminiscent of kinetic constraints in models of glassy
relaxation.  These constraints typically lead to extremely slow kinetics
\cite{JE,SE,MDG,CRRS}, as is also observed in our opinion dynamics model.

The rate equations corresponding to the processes in Eqs.~(\ref{MM}) and
(\ref{MSM}) are
\begin{equation}
\label{RE}
\dot M=-2M^2\qquad \dot S=-M\,S+M^2.
\end{equation}
These give the asymptotic behaviors $M\propto t^{-1}$ and $S\propto
t^{-1/2}$.  Thus an approximate rate equation approach already predicts that
stationary domain walls decay slower than mobile walls.  

An important subtlety in the arrangement of domain walls is that an arbitrary
initial opinion state necessarily leads to an {\em even} number of mobile
walls between each pair of stationary walls.  It is also easy to verify that
domain wall sequences of the form $\ldots M_+M_-M_+\ldots$ cannot arise from
an underlying opinion state.  These topological constraints play a crucial
role in the kinetics.

The exact density of mobile walls can be obtained by again mapping the
constrained spin-1 system onto a spin-1/2 system that is equivalent to the
voter model.  In this mapping, we consider both $+$ and $-$ spins as
comprising the same (non-zero) spin state, while the zero spins comprise the
other state.  With this identification, the reduced model is just the
spin-1/2 ferromagnetic Ising chain with zero-temperature Glauber dynamics and
{\em no} kinetic constraint.  In this reduced model, domain walls $M_+$ and
$M_-$ are indistinguishable and they diffuse and annihilate when upon
colliding.  The density of mobile walls $M(t)=M_+(t)+M_-(t)$ is known exactly
for arbitrary initial conditions from the original Glauber solution
\cite{glauber}.  For initially uncorrelated opinions and if the magnetization
of the spin-1/2 system -- here the difference between the density of non-zero
and zero spins -- equals $m_0$, then \cite{amar}
\begin{equation}
\label{nt}
M(t) = {1-m_0^2\over 2}\,e^{-2t}\left[I_0(2t)+I_1(2t)\right],
\end{equation}  
where $I_k$ is the modified Bessel function of index $k$.  

\begin{figure}[ht] 
 \vspace*{0.cm}\hspace*{-0.1in}
 \includegraphics*[width=0.5\textwidth]{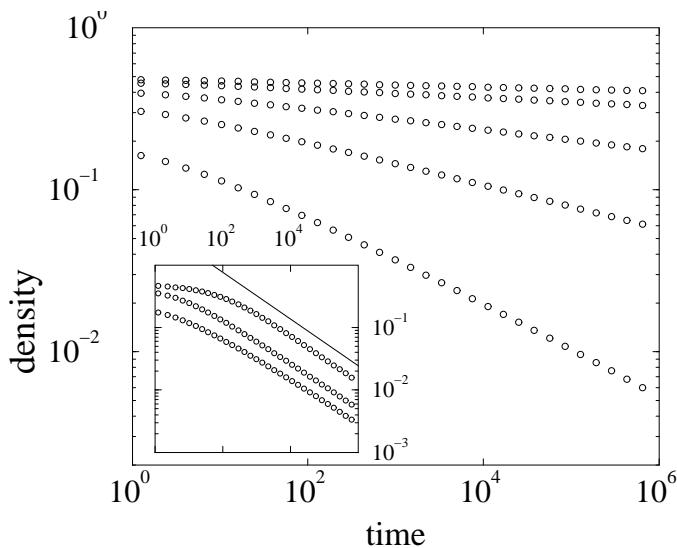}
\caption{Stationary  domain walls density versus time on a double logarithmic 
  scale for the initial conditions $\rho_0= 0.02$, 0.04, 0.10, 0.20, and 0.40
  (top to bottom).  The respective exponent estimates are 0.013, 0.026,
  0.065, 0.13, and 0.29.  Data are based on 100 realizations on a $5\times
  10^5$-site chain.  Inset: Stationary domain wall density for initially
  uncorrelated walls for $\rho_0= 0.02$, 0.10, and 0.40 (top to bottom).  The
  solid line has slope $-3/8$.}
\label{S-density}
\end{figure}

In the spin-1 system, $m_0=\rho_++\rho_--\rho_0$, or $m_0=1-2\rho_0$ by
normalization.  If the initial densities of $+$ and $-$ opinions are equal,
then $M_+(t)=M_-(t)$ and their densities are
\begin{eqnarray}
\label{nt+-}
M_\pm(t)
&=&\rho_0(1-\rho_0)\,e^{-2t}\left[I_0(2t)+I_1(2t)\right]\nonumber \\
&\sim&\rho_0(1-\rho_0)\,(\pi t)^{-1/2}.
\end{eqnarray}  
As expected, the mobile wall density asymptotically decays as $t^{-1/2}$
because of the underlying diffusive dynamics.  However, we find numerically
that the density of stationary domain walls $S(t)$ decays as
\begin{equation}
\label{nts}
S(t)\propto t^{-\psi(\rho_0)},
\end{equation} 
with a {\em non-universal\/} exponent $\psi(\rho_0)$
that goes to zero as $\rho_0\to 0$
(Fig.~\ref{S-density}).

To help understand the mechanism for the slow decay of the stationary domain
wall density, we also simulated a test system with spatially uncorrelated
domain walls.  While such a domain wall state cannot arise from any initial
set of opinions, we can prepare directly an uncorrelated arrangement of
domain walls with prescribed densities.  For any initial condition in this
test system, the stationary wall density decays as $t^{-3/8}$ (inset to
Fig.~\ref{S-density}), consistent with known results on persistence
\cite{per,BKR,stauffer}.  Here persistence refers to the probability that a
given lattice site is not hit by any diffusing domain wall.  For the kinetic
spin-1/2 Ising model, the persistence probability decays as $t^{-\theta}$,
with $\theta=3/8$ \cite{der}, independent of the initial domain wall density,
when the walls are initially uncorrelated.  Thus the topological constraints
imposed on the domain wall arrangement by the initial opinion state appear to
control the dynamics.

These topological constraints lead to the initial-condition dependence of the
amplitude in the mobile wall density (Eq.~(\ref{nt+-})).  This arises because
for $\rho_0\to 0$ the system initially consists of long strings of stationary
walls that are interspersed by pairs of mobile walls, and their survival
probability is proportional to their initial (unit) separation \cite{redner},
leading to the asymptotic density for mobile walls is $M\sim
2\rho_0/\sqrt{\pi t}$ (Eq.~(\ref{nt+-})).  We now exploit this observation to
estimate the density of stationary walls as $\rho_0\to 0$.  Within a
rate-equation approximation, the density of stationary domain walls decays
according to
\begin{equation}
\label{Idot}
\dot S = - k\, M\,S\,.
\end{equation}
While such an equation is generally inapplicable in low spatial dimension, we
can adapt it to one dimension by employing an effective time-dependent
reaction rate $k\sim \sqrt{2/\pi t}$ \cite{BKR,redner}.  This is just the
time-dependent flux to an absorbing point due to a uniform initial background
of diffusing particles; such a rate phenomenologically accounts for effects
of spatial fluctuations in one dimension.  Substituting the asymptotic
expression for $M(t)$ from Eq.~(\ref{nt}) and the reaction rate $k\sim
\sqrt{2/\pi t}$ into this rate equation, we find that the density of
stationary walls decays as $t^{-\psi}$ with
$\psi(\rho_0)=\sqrt{8}\,\rho_0/\pi$ as $\rho_0\to 0$.  It is the amplitude in
the density of mobile domain walls that ultimately causes the slow decay in
the stationary wall density.

\begin{figure}[ht] 
 \vspace*{0.cm}
 \includegraphics*[width=0.4\textwidth]{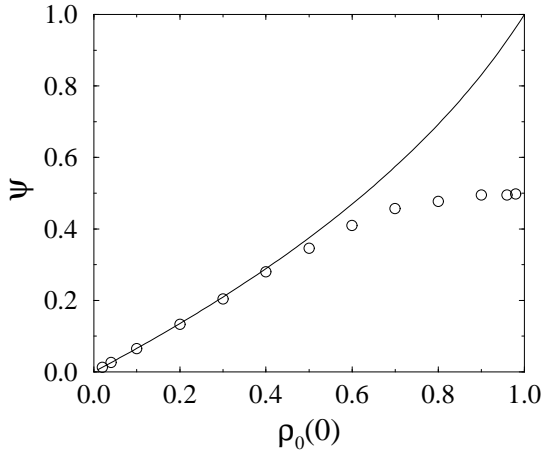}
\caption{Comparison of the exponent $\psi$ from Eq.~(\ref{theta}) and from 
  the simulation data of Fig.~\ref{S-density} (circles).}
\label{exponent}
\end{figure}

A more compelling way to determine $\psi(\rho_0)$ is via persistence in the
$q$-state Potts model.  Because the initial magnetization in the reduced
spin-1/2 system is $m_0=1-2\rho_0$, it has been argued (see e.g. \cite{sire})
that this system should be identified with the $q$-state Potts model with
$m_0=2/q-1$, or $q=(1-\rho_0)^{-1}$.  Using the exact persistence exponent
for the $q$-state Potts model with Glauber kinetics \cite{der}, and
identifying $\psi$ with this persistence exponent, we obtain
\begin{equation}
\label{theta}
\psi(\rho_0)=-{1\over 8}+{2\over \pi^2}\left[\cos^{-1}
\left({1-2\rho_0\over \sqrt{2}}\right)\right]^2,
\end{equation}  
with the limiting behavior $\psi(\rho_0)\to 2\rho_0/\pi$ as $\rho_0\to 0$.
This asymptotics agrees with our numerical results for $\rho_0\alt 0.4$
(Fig.~\ref{exponent}) but then deviates for larger $\rho_0$, where
$\psi(\rho_0)$ must monotonically approach 1/2 as $\rho_0\to 1$.  It should
be noted that in this identification with persistence, we have ignored the
creation of stationary interfaces due to the meeting of mobile domain walls.
This creation process occurs with a rate $(-dS/dt)_{\rm gain}\propto
t^{-3/2}$ and is subdominant for $\psi<1/2$ \cite{FKB}.

An important characteristic of the system is the spatial arrangement of
domain walls.  From our simulations, the mean distances between
nearest-neighbor $MM$ and $MS$ walls both appear to grow as $t^{1/2}$ due to
the diffusive motion of mobile domain walls.  The distributions of these two
distances both obey scaling, with scaling function of the form $ze^{-z^2}$,
where $z=x(t)/\langle x(t)\rangle$ is the scaled separation between walls.
In contrast, the distances between neighboring stationary walls $x_{SS}$ appear
to be characterized by two length scales.  There are large gaps of length of
the order of $t^{1/2}$ that are cleared out by mobile walls before they
annihilate, but there are also many very short distances remaining from the
initial state (Fig.~\ref{dw}).  The corresponding moments $\langle
x_{SS}^k(t)\rangle^{1/k}$ reflect this multiplicity of scales, with $\langle
x_{SS}^k(t)\rangle^{1/k}$ approaching a $t^{1/2}$ growth law as $k\to\infty$,
and growing extremely slowly in time for $k\to 0$.

\begin{figure}[ht] 
  \vspace*{0.cm} \includegraphics*[width=0.4\textwidth]{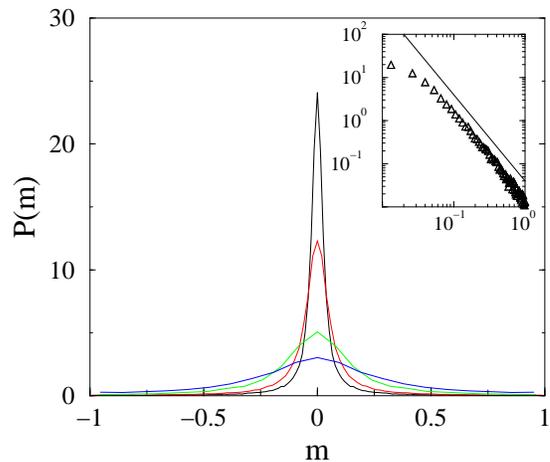}
\caption{Magnetization distribution $P(m)$, ($m=\rho_+- \rho_-$) in the 
  frozen final state for $\rho_0=0.02$ ($10^5$ realizations), 0.04, 0.1, and
  0.2 ($10^4$ realizations) on a $5000$ site linear chain.  Inset: $P(m)$ for
  $\rho_0=0.02$ on a double logarithmic scale to illustrate the $m^{-2}$
  tail.  The straight line has slope $-2$.}
\label{final-m}
\end{figure}

Finally, we quantify the frozen final state by the magnetization distribution
$P(m)$, namely, the density difference between $+$ and $-$ spins.  This
distribution becomes broader as $\rho_0$ increases (Fig.~\ref{final-m}),
reflecting the fact that there is progressively more evolution before the
system ultimately freezes.  For small $\rho_0$, $P(m)$ has a $m^{-2}$ tail.
We may explain this result by considering the evolution of a single pair of
mobile walls.  This pair annihilates at time $t$ with probability density
$\Pi(t)\propto t^{-3/2}$.  The total magnetization of the resulting frozen
state scales as $t^{1/2}$ since the domain wall pair annihilates at a
distance $x\propto t^{1/2}$ from its starting point.  Then from
$P(m)\,dm=\Pi(t)\,dt$, together with $\Pi(t)\propto t^{-3/2}$ and $m\propto
\rho_0 t^{1/2}$, we obtain $P(m)\propto \rho_0\, m^{-2}$.  While the argument
has been formulated in one dimension, we expect this result to apply in all
spatial dimension.

\begin{figure}[ht] 
  \vspace*{0.1cm} \includegraphics*[width=0.45\textwidth]{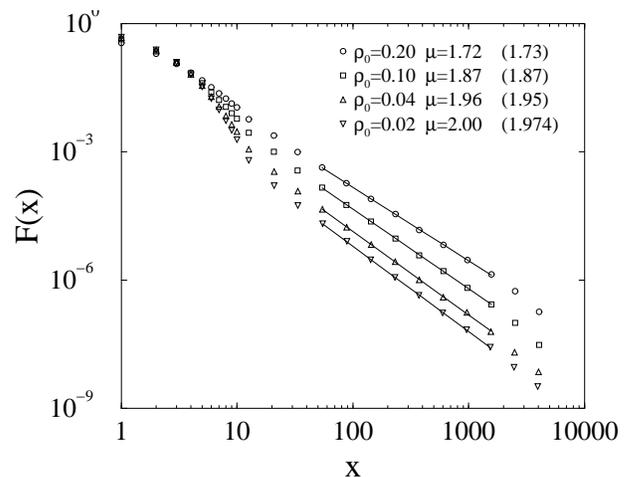}
\caption{Domain length distribution $F(x)$ in the frozen final state for 
  the same data in Fig.~\ref{final-m}.  The data have been binned over a
  small logarithmic range to reduce fluctuations.  Tabulated are the slopes
  from each data set and the expected value $2(1-\psi)$ (parentheses) from
  our simulation result for $\psi$.  }
\label{dist}
\end{figure}

The frozen state is reached when $t=T_f\propto L^2$; this is the time needed
for mobile domain walls to diffuse throughout the system and thus be
eliminated.  At this time, the density of stationary domain walls is of the
order of $S\propto T_f^{-\psi}\propto L^{-2\psi}$.  Thus the average length
of single-opinion domains is $\langle x\rangle\propto L^{2\psi}$.
Numerically we find that the domain length distribution has a power law tail,
$F(x)\propto x^{-\mu}$, with $1<\mu<2$.  The lower bound ensures
normalizability while the upper bound implies that $\langle x\rangle$
diverges as $L\to\infty$.  {}From the above power law form, $\langle
x\rangle=\int dx\,xF(x)\propto L^{2-\mu}$.  This matches our previous
estimate of $\langle x\rangle\sim L^{2\psi}$ when $\mu=2(1-\psi)$.
Figure \ref{dist} shows the length distribution of single-opinion domains in
the frozen state.  Direct estimates of the exponent $\mu$ from this plot are
in good agreement with the exponent relation $\mu=2(1-\psi)$, with $\psi$
obtained from the time dependence of the mobile wall density in
Fig.~\ref{S-density}.

In summary, the constraint that extremists with opposite opinions cannot
influence each other substantially slows down opinion dynamics.  In one
dimension, the density of stationary interfaces between neighboring $+$ and
$-$ spins decays as $t^{-\psi}$, with $\psi(\rho_0)\sim 2\rho_0/\pi$ as
$\rho_0\to 0$.  This slow dynamics is a consequence of the subtle topological
constraints on the arrangement of domain walls, together with the
incompatibility constraint that neighboring $+$ and $-$ spins do not
interact.  

The final opinion outcome depends non-trivially on the initial densities of
the leftist, rightist and centrist opinion states.  With probability $\rho_0$
the final population consists of only centrists, while with probability
$1-\rho_0$ the final population does not contain any centrists.
Analytically, we only know that consensus of centrists is reached with
probability $P_0=\rho_0$; the determination of the complementary final state
probabilities for general initial conditions $P_+$, $P_-$, and $P_{+-}$
remains an open challenge.  Finally, a frozen mixture final state would not
exist if there is a non-zero rate (however small) for opposite extremists to
influence each other directly.  Perhaps this lack of direct influence can
account for the sad phenomenon of the proximity of incompatible populations
in too many locations around the world.

We thank S. Majumdar for helpful correspondence.  We are also grateful to NSF
grant DMR9978902 for partial support of this work.

\end{document}